\documentclass[11pt,twoside]{article}  
\usepackage{adassconf}

\begin{document}   

%
%
%

\paperID{P1-1-23}

%
%
%
%

\title{Structured Query Language for Virtual Observatory}

%
%
%

\author{Yuji Shirasaki, Masatoshi Ohishi, Yoshihiko Mizumoto, 
        Masahiro Tanaka, Satoshi Honda, Masafumi Oe}
\affil{National Astronomical Observatory of Japan, 2-21-1 Osawa, Mitaka City, 
       Tokyo 181-8588, Japan}
\author{Naoki Yasuda}
\affil{Institute for Cosmic Ray Research,  University of Tokyo,
       5-1-5 Kashiwa-no-Ha, Kashiwa City, Chiba 277-8582, Japan}
\author{Yoshifumi Masunaga}
\affil{Ochanomizu University, 2-1-1 0tsuka,Bunkyo-Ku, Tokyo 112-8610, Japan}

%
%

\contact{Yuji Shirasaki}
\email{yuji.shirasaki@nao.ac.jp}

%
%
%
%
%

\paindex{Shirasaki, Y.}
\aindex{Ohishi, M }
\aindex{Mizumoto, Y }
\aindex{Tanaka, M }
\aindex{Honda, S }
\aindex{Oe, M }
\aindex{Yasuda, N }
\aindex{Masunaga, Y }

%
%

\authormark{Shirasaki et al.}

%
%



\begin{abstract}          
Currently two query languages are defined as standards for the 
Virtual Observatory (VO). 
Astronomical Data Query Language (ADQL) is used for catalog data query
and Simple Image Access Protocol (SIAP) is for image data query.
As a result, when we query each data service, we need to know in advance 
which language is supported and then construct a query language
accordingly.
The construct of SIAP is simple, but they have a limited capability.
For example, there is no way to specify multiple regions in one query, 
and it is difficult to specify complex query conditions.
%
%
In this paper, we propose a unified query language for any kind
of astronomical database on the  basis of SQL99.
SQL is a query language optimized for a table data, so to apply the SQL
to the image and spectrum data set, the data structure need to be mapped
to a table like structure.
%
%
We present specification of this query language and an example of the
architecture for the database system.
\end{abstract}

%
%

\section{Introduction}

At present, two kinds of data query language are defined as standards in
the Virtual Observatory.
One is the parameter query, which is used for image data search and
called as ``Simple Image Access Protocol'' or SIAP.
Search criteria are specified by a set of ``key'' and ``value'' pairs.
Another one is a structured query language, which is used for catalog 
data query and called as ``Astronomical Data Query Language'' or ADQL
(Yasuda et al. 2004).
ADQL can specify more complex search criteria than the SIAP does, 
also has an ability to join multiple tables, and can select values
derived from the DB columns.
So it is worthwhile to adapt the ADQL to perform an image query.
The high flexibility of the ADQL syntax, however, raise difficulties to
develop an ADQL compliant data service.
It is known that incompatibility among various DBMS products is
due to the complexity of the SQL specification.
The success of the Virtual Observatory project depends on uniformity
of interfaces of all the astronomical data services, so we need to avoid
the complexity and make it simple and easy to implement.

\section{Syntax Specification}

In order to realize the interoperability among the distributed data
archives, we need to define a standard query language.
As the standard query language must be supported by all the data services,
it's specification should be simple and clear for easy implementation.
On the other hand, some data service require more sophisticated query
syntax to allow users to specify more efficient search criteria.
Thus we defined basic syntax which must be supported by all the data
services and allowed several enhancements on the basic syntax.
\subsection{Basic syntax}
The following restrictions are applied to the ``select'' statement of
SQL99 specification.
\begin{enumerate}
\item Only a column name or ``*'' is specified in the selection list, so
      an algebraic expression and a function are not allowed.
      So the role of the ``select'' clause is just to specify the
      columns you want to get.
      The calculation of the derivative from those columns should be
      done on a portal server or by users themselves.
\item Only one table is specified in the ``from'' clause, so any type of
      table join syntax is not allowed.
      The join of the tables should be done on a portal server or by
      users themselves.
\item Allowed comparison operators and predicates are \verb|=|, \verb|<|,
      \verb|>|, \verb|>=|, \verb|<=|, \verb|<>|, \verb|within|,
      \verb|contains|, \verb|overlaps|, \verb|LIKE| and \verb|BETWEEN|,
      and any other operators are not allowed.
      The operators \verb|within|, \verb|contains| and \verb|overlaps|
      are not specified in SQL99, but are introduced to specify a
      region on the sky as such search criterion is a fundamental one
      for the astronomical data service.
      These three operators compare two values of geometry data type,
      which is described in the next subsection.
\item Functions \verb|point()|, \verb|circle()| and \verb|box()| are
      introduced to express a region on the sky, and \verb|distance()|
      is also introduced to describe the proximity of the two coordinates.
\item Allowed logical operators are \verb|AND| and \verb|NOT|.
      \verb|OR| is not allowed as the mixture of \verb|AND| and
      \verb|OR| makes the search condition too complex.
\item Table and column may have an alias name.
\end{enumerate}
\subsection{Geometry Data Type}

The geometry data type is introduced to represent a point or a region in
the sky.
A point is always expressed by two coordinate values, and in most of
the cases a single coordinate value is meaningless and only the pair
have a physical meaning.
So it is natural to have a column which has a pair of values to express
the coordinate of celestial object, then it becomes simple to describe
the search criterion.
For an example, if the coordinate values are prepared in two separate
columns, the search condition is expressed as follows:
\verb|point(t1.ra, t1.dec) within box((t2.ra, t2.dec), 1.0, 1.0)|.
On the other hand, if the coordinate values are prepared in one single
column, the same statement can be expressed shortly as follows:
\verb|t1.point within t2.region|.
We defined ``Point'', ``Circle'' and ``Box'' data types.
Expressions of these data types in SQL are summarized in 
Table~\ref{tbl:geo_data_type}.
\begin{deluxetable}{cl}
   \small
   \tablecaption{Expressions of the geometry data type value}
   \tablehead{\colhead{Data Type} & \colhead{Examples of Expression}}
   \startdata
Point & \verb|Point(23., +10., 'FK5')|, \verb|Point(23., +10.)|\verb|, (23., +10.)|\\ 
Box   & \verb|Box((23.0, +10.0), 1, 1)|, \verb|((23.0, +10.0), 1, 1)|\\ 
Circle & \verb|Circle((23.0, +10.0), 1)|, \verb|((23.0, +10.0), 1)|\\
   \enddata
   \label{tbl:geo_data_type}
\end{deluxetable}
\subsection{Example SQLs for Basic Syntax}

The next SQL shows an example of catalog data query.
\begin{verbatim}
  Select g.ra, g.dec, g.mag_r
  From   galaxy as g
  Where  Point(g.ra, g.dec) within Circle((24.3, +5.0), 1.0) 
         and g.mag_r < 24
\end{verbatim}
This query describes ``from a table named galaxy select right ascension,
declination, and R-band magnitude of celestial objects located in the
circle whose center coordinate is (ra, dec) = (24.3 deg, +5.0 deg) and
radius is 1.0 deg''.
In the ``galaxy'' table, coordinates of the objects are prepared in the
two separate columns, ``ra'' and ``dec'', so  Point()  function is
required to describe the search region.
In the ``galaxy2'' table of the next SQL sample, the coordinates are
prepared in a single column ``point''.
\begin{verbatim}
  Select g.point, g.mag_r
  From   galaxy2 as g
  Where  g.point within Circle((24.3, +5.0), 1.0) and g.mag_r < 24
\end{verbatim}
The next SQL shows an example of an image data query.
\begin{verbatim}
  Select img.filter, img.imageURL
  From   imageData as img
  Where  img.region = BOX((24.3, +5.0), 0.1, 0.1)
\end{verbatim}
This equivalents to the next URL-based SIAP.
\begin{verbatim}
  http://jvo.nao.ac.jp/image?POS=24.3,5.0&SIZE=0.1&FORMAT=VOTABLE
\end{verbatim}
In the above SQL, the column \verb|region| is used as a parameter to specify
the region of interest, and the access URL of the cut-out image is
returned as a data of \verb|imageURL| column.
Those columns are actually not present in the relational table on the
DBMS, but user can specify as if it virtually exists.
So these columns are called as virtual columns and its table is called
as a virtual table or a view in the RDB terminology.
\subsection{Syntax Enhancement}
The following enhancements may be applied as optional features.
\begin{enumerate}
\item Multiple tables enhancement: 
      Multiple tables can be queried with a single SQL. 
      External tables provided in a form of VOTable must be treated in
      the same manner as the original tables of the data service.
      The external table is specified as
      \verb|EXT::<fileNumber>.<resourceName>.<tableName>|.
      Join predicate and sub-query are not mandatory.
\item Unit support enhancement: 
      Any numeric value may be followed by a unit. 
      Unit conversion must be carried out by the data service.
\item Algebraic expression enhancement: An algebraic expression can be
      specified in the selection list and search criteria in the ``where''
      clause.
\item Logical operator \verb|OR| enhancement: 
      Support for mixture of \verb|AND|, \verb|NOT| and \verb|OR| in the
      ``where'' clause.
\item Object data type enhancement: 
      A column may have a structured data and access methods.
\item Use of identifier for specifying a table name (portal): 
      A table name is expressed by an identifier of dot notation to
      specify the table uniquely in the VO.
      A dot character in the identifier must be escaped by a back slash.
      For an example, \verb|naoj:sxds.v0\.1.galaxy|) where \verb|naoj|
      is an authority name, \verb|/sxds/v0.1| is a catalog resource
      path name and \verb|galaxy| is a table resource name.
      This is a feature dedicated to a portal data service.
\item UCD (portal): 
      A UCD (Derriere et al. 2004) can be used in place of a column
      name. 
      This is a feature dedicated to a portal data service. 
      The portal service searches data resources which have UCD
      specified in the SQL and translate to the column name using the
      column meta data collected from the data services.
\item Omission of ``From'' part (portal): 
      This is a feature dedicated to a portal data service.
      The portal searches the data resources from the registry according
      to the query condition.
      In this case, UCD must be used for describing columns
      and a query condition.
\end{enumerate}

\section{JVO Skynode Architecture}
\begin{figure}
   \epsscale{0.8}
   \plotone{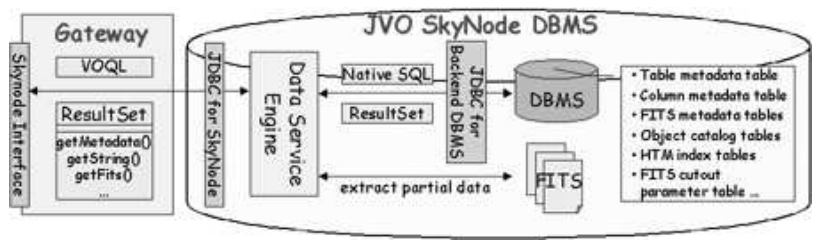}
   \caption{Architecture of JVO Skynode designed to be compatible with
    the proposed VOQL.}
   \label{fig:skynode_arc}
\end{figure}
\begin{figure}
   \epsscale{0.8}
   \plotone{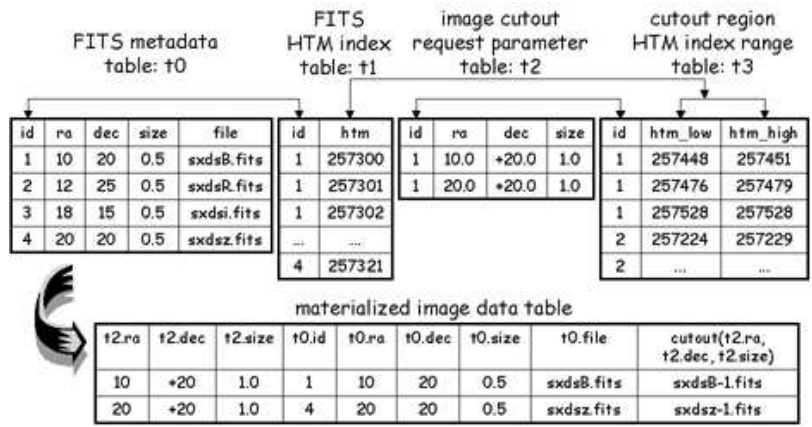}
   \caption{An example to materialize the virtual image table.}
   \label{fig:virtual_image_table}
\end{figure}

We have designed an architecture to implement a data service compatible
with the proposed query language (Figure~\ref{fig:skynode_arc}).
The ``gateway'' in the figure provides interfaces to access the
skynode (Yasuda et al. 2004) services.
%
%
The received query is transfered to the JVO SkyNode DBMS, and the result
is returned in a form of a \verb|ResultSet| Java object.
The access interface is provided by JDBC which is a standard interface
to access the DBMS in the Java environment.
Celestial object catalog is stored in a relational data base (RDB)
system, such as PostgreSQL, MySQL, Oracl, and so on, and FITS files are
managed in the unix files system and its meta data is stored in the RDB.
HTM index (Kunszt et al. 2000) is used to perform a fast data search.
Query condition related to the search region on the sky is converted to
a condition on the HTM index ranges at the ``Data Service Engine'', 
and then the modified SQL is submitted to the DBMS.
In the case of image data query, it is necessary to materialize the
virtual image table as follows:
First, parameters specifying the image cut out regions are stored in a
table t2 (Figure~\ref{fig:virtual_image_table}) and the corresponding
HTM index range table t3 is also created.
Then FITS meta data table t0 is joined with table t2 with intervention
of the HTM index table t1 and t3, then the virtual image table is
materialized as shown in the bottom of the figure.

\acknowledgments

This work was supported by the JSPS Core-to-Core Program and
Grant-in-aid  Information Science  (15017289 and 16016292) carried out
by the Ministry of Education, Culture, Sports, Science and Technology of
Japan.

\end{document}